# Strain-induced Isostructural and Magnetic Phase Transitions in Monolayer MoN$_2$

Yao Wang[†‡], Shan-Shan Wang[§], Yunhao Lu[†‡*], Jianzhong Jiang[†‡*], Shengyuan A. Yang[§*]

[†]International Center for New-Structured Materials (ICNSM) and School of Materials Science and Engineering, Zhejiang University, Hangzhou 310027, China
[‡]State Key Laboratory of Silicon Materials, Zhejiang University, Hangzhou 310027, China
[§]Research Laboratory for Quantum Materials, Singapore University of Technology and Design, Singapore 487372, Singapore

**ABSTRACT:** The change of bonding status, typically occurring only in chemical processes, could dramatically alter the material properties. Here, we show that a tunable breaking and forming of a diatomic bond can be achieved through physical means, i.e., by a moderate biaxial strain, in the newly discovered MoN$_2$ two-dimensional (2D) material. Based on first-principles calculations, we predict that as the lattice parameter is increased under strain, there exists an isostructural phase transition at which the N-N distance has a sudden drop, corresponding to the transition from a N-N nonbonding state to a N-N single bond state. Remarkably, the bonding change also induces a magnetic phase transition, during which the magnetic moments transfer from the N(2$p$) sublattice to the Mo(4$d$) sublattice, meanwhile the type of magnetic coupling is changed from ferromagnetic to anti-ferromagnetic. We provide a physical picture for understanding these striking effects. The discovery is not only of great scientific interest in exploring unusual phase transitions in low-dimensional systems, but it also reveals the great potential of the 2D MoN$_2$ material in the nanoscale mechanical, electronic, and spintronic applications.

Breaking and forming of chemical bonds are at the core of chemical reactions, typically of energy scales much larger than those associated with physical processes, hence seem unlikely to be controllable by physical means in a reversible manner. The obstacle may be somewhat alleviated when considered in a solid-state environment, due to the strong coupling between lattice geometry and electronic configuration. Indeed, striking observations have been made in the past on a few bulk crystalline materials, particularly for a family of rare-earth transition metal phosphides,[1-3] that a pressure-induced isostructural phase transition could occur, during which the bonding status between two P atomic sites changes completely, with dramatic influence on the electronic properties. However, such instances are still rare and the transition usually requires a very high pressure that is not easily accessible.[4-6]

Starting from the discovery of graphene,[7] the field of two-dimensional (2D) materials has undergone rapid development in the past decade. Many new kinds of 2D materials have been proposed and fabricated, which exhibit a vast range of novel material properties, not seen in the usual 3D bulk materials.[8-10] In particular, 2D materials of atomic thickness typically have great mechanical flexibility, e.g., graphene, $MoS_2$, and phosphorene can sustain lattice strains above 25%.[11-15] And for 2D materials, strain application is also more readily controlled in experiment, e.g., by stretching or bending of a flexible substrate on which the 2D layer is attached.[16, 17] These observations lead us to speculate that 2D materials might offer a promising platform to search for the fascinating bonding-nonbonding phase transitions induced by strain.

Meanwhile, one current challenge in the 2D materials' research is to realize intrinsic magnetic 2D materials and to design feasible methods to control the magnetic ordering. Attempts have been made, mostly by introducing magnetic dopants, e.g., with transition metal doping for graphene and dichalcogenides 2D materials.[18-21] However, the achieved magnetism is usually weak, and more crucially, it is difficult to control the distribution of the extrinsically-introduced defects in experiment, which may strongly scatter the carriers and deteriorate the material's transport properties. Therefore, it is much desired to have an intrinsic magnetic 2D material and at the same time to have its magnetic properties easily tunable by physical means.

Remarkably enough, in this work, we show that the above-mentioned intriguing effects and properties can be manifested in a single 2D material—the monolayer $MoN_2$. As a new layered molybdenum dinitride material, the bulk single-crystal $MoN_2$ was successfully synthesized in recent experiment.[22] Subsequent study has shown that a monolayer $MoN_2$ could be realized and its ground state has a strong ferromagnetic ordering with a Curie temperature above 420 K.[23] Here, based on first-principles calculations, we discover that the monolayer $MoN_2$ actually exhibits two stable isostructural phases with extremely dissimilar N-N distances along the perpendicular direction: the $\alpha$-phase of a large N-N distance ($d_{N-N} \cong 2.20$ Å), corresponding to the ground state structure discussed in the previous study;[23] and a new $\beta$-phase of a much shorter N-N distance ($d_{N-N} \cong 1.54$ Å). We show that the key difference between the two phases is on the N-N bonding status: the $\alpha$ phase is with a nonbonding N-N state; whereas the $\beta$-phase is with a N-N single bond state. Importantly, the transition between the two phases can be easily induced by applying a small biaxial strain ~5% (referenced to the unstrained structure in the $\alpha$ phase), at which $d_{N-N}$ has an abrupt change corresponding to the N-N bonding-nonbonding transition. Furthermore, we find that the drastic change in bonding at the isostructural phase transition strongly affects the magnetic ordering: the $\alpha$-phase has a ferromagnetic (FM) ground state with magnetic moments mainly on the N(2$p$) sublattice; upon the $\alpha$ to $\beta$ phase transition, the moments disappear on the N(2$p$) sublattice, meanwhile the moment on Mo(4$d$) sublattice is greatly enhanced and forms an anti-ferromagnetic (AFM) ordering. Therefore, our discovery reveals a remarkable phenomenon of tunable bond forming and breaking in a solid-state material, and to our best knowledge, this is the *first* report of a simultaneous isostructural and magnetic phase transition realized in a 2D material. Our findings will pave the way for the many promising electronic, mechanic, and spintronic applications based on this novel 2D material.

Our first-principles calculations are based on the density functional theory (DFT) using the Perdew-Burke-Ernzerhof (PBE)[24] realization of the generalized gradient approximation (GGA) for the exchange-correlation, as implemented in the Vienna Ab-initio Simulation Package (VASP).[25] The projector augmented wave (PAW)[26] method is employed to model the ionic potentials. Kinetic energy cutoff is set above 500 eV for all calculations. A vacuum space of 25 Å was introduced to avoid interactions between images. The structure is fully optimized with respect to the ionic positions until the forces on all atoms are less than 0.01 eV/Å. The Monkhorst-Pack *k*-point sampling is used for the Brillouin zone integration: $35 \times 35 \times 1$ for relaxation, and $45 \times 45 \times 1$ for self-consistent calculations. As the transition metal *d* orbitals may have important correlation effects, we also validate our results by using the DFT+*U* method following the approach of Dudarev *et al.*[27] Several on-site Hubbard *U* parameters (*U*=3, 4, 5 eV) are tested for Mo(4*d*) orbitals, which yield almost the same results as that from GGA calculations. Our main results are further confirmed by calculations with hybrid functional method (HSE06). Hence in the following discussion, we will focus on the GGA results.

The 3D bulk $MoN_2$ adopts a rhombohedral *R3m* space group and has a layered structure. The interaction between the $MoN_2$ layers is weak, with a low exfoliation energy, hence it has been proposed that the monolayer $MoN_2$ can be exfoliated using similar methods as for other 2D materials.[23] The derived monolayer $MoN_2$ has the hexagonal lattice symmetry with trigonal prismatic coordination, as shown in Figs. 1(a) and 1(b). The structure has the same symmetry as the 1H-MoS$_2$, with two N atomic layers sandwiching a single Mo layer in a mirror-symmetric manner. In the $\alpha$-phase (which is the one studied in the previous work), the fully-relaxed structure that we obtain has a lattice parameter of $a = 2.98(4)$ Å, in good agreement with the previous work.[23] The N-Mo bond length is 2.04(3) Å, smaller than that for the S-Mo distance (~2.41 Å) in MoS$_2$ due to the smaller ionic radius of N. We will pay particular attention to the N-N distance $d_{\text{N-N}}$ between the two N sites along the vertical direction (see Fig.1(b)). In the $\alpha$-phase, $d_{\text{N-N}} = 2.19(7)$ Å, which is far beyond the typical N-N bond length (~1.45 Å for a single bond),[28] resulting in a negligible bonding interaction, hence the $\alpha$-phase has a non-bonding N-N state. Our calculation also confirms that the ground state in $\alpha$-phase takes a FM ordering. As shown in Fig.1(c), the energy of the FM configuration (marked by A) is lower than that of the non-spin polarized configuration by about 0.2 eV per formula unit. The magnetic moment is mainly from the N(2$p$) sublattice (~0.42 $\mu_B$), whereas the Mo(4$d$) moment is much smaller (~0.13 $\mu_B$).

Starting from the $\alpha$-phase, when we apply an in-plane biaxial strain, the system energy increases in a quadratic way, as for typical elastic deformations (see Fig.1(c)). One observes that the trend terminates at $a$~3.13 Å (i.e., ~5% strain), where the slope of the energy curve (for both FM and non-spin) has an abrupt change of sign, forming a cusp-like shape. Further increasing the lattice parameter lowers the energy, and importantly, there exists a second (local) energy minimum, after which the energy curve increase again up to very large lattice parameters ($>$ 3.6 Å). One notes that in $\alpha$-phase (below the transition at $a$~3.13 Å), the FM configuration always has lower energy than the non-spin polarized one. However, just above 3.13 Å, the two energy curves almost coincide with negligible energy difference, implying that the FM ordering is no longer favored above the transition. Is it possible to have a ground state with another type of magnetic ordering? In the calculation, we further consider the striped AFM configuration (see Fig.1(a)). As shown in Fig.1(c), in the region above $a \sim 3.13$ Å, the AFM configuration indeed has lower energy compared with both FM and non-spin polarized configurations. The energy minimum occurs at $a = 3.30$ Å (marked as B) with an energy only slightly higher than that of state A. Meanwhile, we find that the magnetic moment for this state is almost entirely located on the Mo sites with a magnitude ~0.83 $\mu_B$.

The discontinuity in the slope of the energy curve at $a \sim 3.13$ Å signals a first-order phase transition. As discussed, in the transition from $\alpha$-phase ($a < 3.13$ Å) to $\beta$-phase ($a > 3.13$ Å), the magnetic ordering changes from FM to AFM, which corresponds to a magnetic phase transition. However, as observed from Fig.1(c), the cusp-like shape is present for each of the three energy curves, also including the non-spin polarized one. This suggests that the change in the magnetic ordering is not the dominant driving force of the phase transition, rather, it is an induced one. In search of the fundamental driving mechanism of the phase transition, we notice that across the transition, there is no change in the crystalline symmetry: the structures in $\alpha$- and $\beta$-phases have the same space group, but the N-N distance $d_{\text{N-N}}$ has a sudden change. As shown in Fig.1(d), at the transition point $d_{\text{N-N}}$ drops by almost 19% from ~2.1 Å to ~1.7 Å. The difference in the N-N distance is even larger between the two stable structures A and B: $d_{\text{N-N}} \cong$ 2.20 Å for A whereas $d_{\text{N-N}} \cong 1.54$ Å for B. The N-N distance in $\beta$-phase becomes much closer to the typical N-N single bond length (~1.45 Å), which implies the possible transition from N-N nonbonding state to the N-N single bond state during the $\alpha \to \beta$ transition.

To verify the above picture, we calculate the electron localization function (ELF) for states A and B in the two phases. The ELF indicates the degree of electron localization and is useful for the analysis of bonding characters. In Fig.2(a), we plot the ELF map in the vertical plane containing the two N atoms and one Mo atom. In $\alpha$-phase, it is clear that there is no bonding between the two N ions. In contrast, in $\beta$-phase, a covalent bonding character appears between the two N sites, indicating the formation of a N-N bond. This transition is also evident in the change of the charge density distribution. In Fig.2(b), we show the deformation charge density for the two states, defined as the total charge density with the density of isolated atoms subtracted. Focusing on the space between the two N sites, one clearly observes that there is charge accumulation in $\beta$-phase which is absent in $\alpha$-phase. Hence the comparisons in Fig.2 indeed show that the N-N bonding status changes from non-bonding to bonding across the $\alpha \rightarrow \beta$ phase transition. Similarly, if we start with $\beta$-phase and apply a compressive strain, the system can be driven to $\alpha$-phase with a N-N bonding to non-bonding transition at $a \sim 3.13$ Å (~5% compressive strain as referenced to state B). These calculations confirm our previous speculation that the N-N bonding change is the underlying driving mechanism that characterizes the phase transition. Since the symmetry of the structure (the space group) remains unchanged, this transition corresponds to an isostructural phase transition.

We further verify that the two energy extremal structures A and B in the two phases are dynamically stable by calculating their phonon spectra. As shown in Fig.3, the absence of imaginary frequencies in the whole Brillouin zone demonstrates the dynamical stability of both structures. For small lattice deformations around the two equilibrium states A and B, the elastic properties of 2D materials can be characterized by the in-plane stiffness constant, defined as $C = (1/S_0)(\partial^2 E_S/\partial \varepsilon^2)$, where $S_0$ is the corresponding equilibrium area, the strain energy $E_S$ is the energy difference between the respective strained and unstrained systems, and $\varepsilon$ is the in-plane uniaxial stain. Through calculation, we find that the stiffness constant is 195 N/m for A and 108 N/m for B. These values are comparable to that of $MoS_2$ (~140 N/m),[29] and are much smaller than that of graphene (~340 ± 40 N/m)[11] and BN (~267 N/m),[30] indicating that the material is softer, which would facilitate the application of external strain.

The bonding change in the isostructural phase transition is expected to severely affect the electronic properties. As reflected in the electronic band structures for A (FM) and B (AFM) states (see Fig. 4(a, b)), the system remains metallic across the transition, but the band structure changes a lot. Besides the difference in the spin-splitting between FM and AFM orderings, the band dispersions around the Fermi level are also completely different between the two states. Projecting the density of states (DOS) onto each atomic orbitals, we find that the states around the Fermi level are mainly from the N-$p_z$ orbital and the Mo-$d_z^2$ orbital (Supporting Information). Furthermore, by integrating the projected density of states (PDOS) below the Fermi level, we note that the total occupation number of the Mo orbitals are almost unchanged between A and B states with an approximate valence of +4.0, which complies with the formal valence change associated with N-N single bond formation ($2N^{2-} \rightarrow N_2^{4-}$). Our calculation result is consistent with this analysis (Fig.5): in $\alpha$-phase, the N-$p_z$ orbital is peaked around the Fermi level and is only partially filled, providing the magnetic moment in the FM state; whereas in $\beta$-phase, the N(2$p$) orbitals are almost fully saturated due to the N-N bonding, therefore the moment at the N site is suppressed.

The suppression of FM ordering in the $\alpha \rightarrow \beta$ phase transition can also be understood in the electronic band picture, in which the spontaneous FM ordering occurs when the relative gain in exchange interaction is larger than the loss in kinetic energy. This leads to the Stoner criterion that the ferromagnetism is realized when $JD(E_F) > 1$,[31, 32] where $J$ is the strength of the exchange interaction and $D(E_F)$ is the (non-spin polarized) DOS at the Fermi level. From Fig.5(a), one observes that in $\alpha$-phase, the strong peak (mainly from the N-$p_z$ orbital) gives a large contribution to the DOS at Fermi level. The narrow band width indicates that these states are quite localized, hence could favor a strong exchange interaction. Similar scenario has been previously discussed in the so-called $d^0$ FM semiconductors.[33, 34] The exchange coupling strength $J$ can be estimated from the spin-splitting in the FM state. From our calculation, we find that the Stoner parameter $JD(E_F)$ is far greater than 1 in $\alpha$-phase (~7.5 for state A), indicating a strong FM instability in this phase. However, at the $\alpha \rightarrow \beta$ transition, $D(E_F)$ has a drastic drop. Due to the change of bonding, the N(2p) contribution is almost depleted at the Fermi level (see Fig.5(b)). Even if we use the same exchange coupling strength (which is surely an overestimation), the Stoner parameter is decreased to ~1 just above the transition, indicating that the spontaneous FM ordering would be suppressed.

In $\beta$-phase, the metallic character is greatly weakened, as reflected in the drop of $D(E_F)$. This is also intuitive since the atoms are further separated (except for the decrease in $d_{N-N}$). The DOS near Fermi level is now dominated by the partially filled Mo-$d_z^2$ orbital component. The electrons in the Mo(4d) orbitals, although typically less localized as those in 3d transition metal elements, can only couple through the saturated N(2p) orbitals, especially the N-$p_z$ orbital in the N-N single bond which is close to the Fermi energy. The scenario is similar to the super-exchange process in typical transition metal oxides.[35] Here the Mo-N-Mo bond angle differs from 90° or 180°, hence the empirical Goodenough-Kanamori-Anderson rules[36-39] cannot directly apply. Nevertheless, as we mentioned, the intermediate N(2p) orbital involved in the exchange process is mainly the N-$p_z$ orbital in the N-N single bond without orbital degeneracy, so exchange interaction between the moments on Mo sites would generally favor an AFM configuration.[39]

Before concluding, several remarks are in order. First, we stress that to our best knowledge, this is the first time that a tunable isostructural bonding-nonbonding phase transition is discovered in a 2D material. It is also the first to realize a strain-induced FM to AFM magnetic phase transition for a 2D material. It is very interesting that the magnetism is manifested by two different atomic sublattices for the two different phases.

In 3D bulk materials, as mentioned in the introduction, isostructural phase transitions were observed in the past in a family of rare-earth transition metal phosphides, with $EuCo_2P_2$ as a representative. In that case, the applied pressure can induced a transition at which P-P transforms from nonbonding to bonding state, and there is also a change in magnetic properties: the magnetism transfers from the Eu(4$f$) sublattice to the Co(3$d$) sublattice.[1-3] The most salient difference in our case is the reduced dimensionality: the monolayer $MoN_2$ is a 2D material with atomic thickness. Because of this, the strain needed to induce the transition is small and readily achievable. In comparison, the transition for $EuCo_2P_2$ requires a high pressure ~3.1 GPa. The magnetism transfer in our case is also different: it is between 2$p$ and 4$d$ orbitals. Moreover, the Curie temperature is predicted to be above 420 K for $MoN_2$ in the FM $\alpha$-phase,[23] which is high enough to realize many practical applications, whereas the magnetic critical temperatures for $EuCo_2P_2$ are much lower (~66.5 K for the low-pressure phase and ~260 K for the high-pressure phase).

Secondly, in our calculation, we have considered a particular AFM configuration for the $\beta$-phase, the striped AFM ordering. There could be other types of AFM ordering to compete for the ground state. Particularly, due to the geometrical frustration inherent to the hexagonal lattice, it is difficult to find the exact ground state in calculation, and some even interesting configurations such as the spin spiral state may also occur. A detailed study on this problem is beyond the scope of the current work. Nevertheless, our results do predict the suppression of FM and the formation of certain AFM ordering in the $\alpha \rightarrow \beta$ phase transition. This will lead to pronounced signals in the magnetic measurement (e.g., using SQUID), and magneto-optic or magneto-transport measurements. The magnetic configuration can also be directly probed by neutron scattering or Mössbauer effect experiments.

Thirdly, different types of strain can be applied to 2D materials. As we have demonstrated, the isostructural phase transition with the change of bonding can be induced by the biaxial strain. We also performed the calculation with uniaxial strains, and no signature of the bonding change or the magnetic ordering change is observed. This can be understood by noting that for uniaxial strains, while stretched along one direction, the system can still relax (contract) along the other direction, which makes the change of N-N distance relatively small. In contrast, for biaxial strains, both in-plane directions are constrained, hence the N-N distance can be changed more effectively, leading to the bonding change for the phase transition.

Finally, like $MoS_2$, the three atomic layers X-Mo-X (X stands for N or S) may also take an ABC-type stacking sequence, known as the 1T structure.[40] We have checked that for both 1T-$MoN_2$ and 1T-$MoS_2$, there is no analogous strain induced phase transitions, indicating the importance of lattice geometry in the emergence of isostructural phase transitions.

In conclusion, a tunable bond forming and breaking driven isostructural phase transition has been proposed in the newly discovered $MoN_2$ 2D material based on first-principles calculations. Besides the $\alpha$-phase with a N-N nonbonding state and a FM ordering, we identify a second $\beta$-phase with a N-N single bond state and a AFM ordering. We show that the transition between the two phases can be readily achieved by applying a biaxial strain. At the phase transition, the N-N distance has a sudden change associated with the bond forming or breaking, while the crystalline symmetry is preserved. The magnetic structure is totally transformed during the transition, with a transfer of magnetism from the N(2$p$) sublattice to the Mo(4$d$) sublattice and the magnetic coupling transformed from FM to AFM. We have provided physical pictures for understanding these remarkable phenomena. The predicted effects are expected to offer many promising mechanical, electronic, and spintronic applications based on this 2D material. And as the *first* 2D material ever reported to host these highly nontrivial properties, it will surely inspire further researches to explore similar effects in many other novel materials in the near future.

# REFERENCES

(1) Huhnt, C.; Schlabitz, W.; Wurth, A.; Mewis, A.; Reehuis, M. *Phys. Rev. B: Condens. Matter Mater. Phys.* 1997, 56, (21), 13796-13804.

(2) Abd-Elmeguid, B.; Micklitz, H.; Sanchez, J. P.; Vulliet, P.; Johrendt, D. *Phys. Rev. B: Condens. Matter Mater. Phys.* 2001, 63, (10), 100102.

(3) Chefki, M.; Abd-Elmeguid, M. M.; Micklitz, H.; Huhnt, C.; Schlabitz, W.; Reehuis, M.; Jeitschko, W. *Phys. Rev. Lett.* 1998, 80, (4), 802-805.

(4) Liu, Q.; Yu, X.; Wang, X.; Deng, Z.; Lv, Y.; Zhu, J.; Zhang, S.; Liu, H.; Yang, W.; Wang, L.; Mao, H.; Shen, G.; Lu, Z.; Ren, Y.; Chen, Z.; Lin, Z.; Zhao, Y.; Jin, C. *J. Am. Chem. Soc.* 2011, 133, (20), 7892-7896.

(5) Zhao, J.; Xu, L.; Liu, Y.; Yu, Z.; Li, C.; Wang, Y.; Liu, Z. *J. Phys. Chem. C* 2015, 119, (49), 27657-27665.

(6) Tan, X.; Fabbris, G.; Haskel, D.; Yaroslavtsev, A. A.; Cao, H.; Thompson, C. M.; Kovnir, K.; Menushenkov, A. P.; Chernikov, R. V.; Garlea, V. O.; Shatruk, M. *J. Am. Chem. Soc.* 2016, 138, (8), 2724-2731.

(7) Novoselov, K. S.; Geim, A. K.; Morozov, S. V.; Jiang, D.; Zhang, Y.; Dubonos, S. V.; Grigorieva, I. V.; Firsov, A. A. *Science* 2004, 306, (5696), 666-669.

(8) Xu, M.; Liang, T.; Shi, M.; Chen, H. *Chem. Rev* 2013, 113, (5), 3766-3798.

(9) Butler, S. Z.; Hollen, S. M.; Cao, L.; Cui, Y.; Gupta, J. A.; Gutierrez, H. R.; Heinz, T. F.; Hong, S. S.; Huang, J.; Ismach, A. F.; Johnston-Halperin, E.; Kuno, M.; Plashnitsa, V. V.; Robinson, R. D.; Ruoff, R. S.; Salahuddin, S.; Shan, J.; Shi, L.; Spencer, M. G.; Terrones, M.; Windl, W.; Goldberger, J. E. *Acs Nano* 2013, 7, (4), 2898-2926.

(10) Xu, X.; Yao, W.; Xiao, D.; Heinz, T. F. *Nat. Phys.* 2014, 10, (5), 343-350.

(11) Lee, C.; Wei, X.; Kysar, J. W.; Hone, J. *Science* 2008, 321, (5887), 385-388.

(12) Kim, K. S.; Zhao, Y.; Jang, H.; Lee, S. Y.; Kim, J. M.; Kim, K. S.; Ahn, J.; Kim, P.; Choi, J.; Hong, B. H. *Nature* 2009, 457, (7230), 706-710.

(13) Bertolazzi, S.; Brivio, J.; Kis, A. *Acs Nano* 2011, 5, (12), 9703-9709.

(14) Castellanos-Gomez, A.; Poot, M.; Steele, G. A.; van der Zant, H. S. J.; Agrait, N.; Rubio-Bollinger, G. *Nanoscale Res Lett.* 2012, 7, (233), 1-4.

(15) Peng, X.; Wei, Q.; Copple, A. *Phys. Rev. B: Condens. Matter Mater. Phys.* 2014, 90, (8) ,085402.

(16) Akinwande, D.; Petrone, N.; Hone, J. *Nat. Commun.* 2014, 5, 5678.

(17) Chang, H.; Yang, S.; Lee, J.; Tao, L.; Hwang, W.; Jena, D.; Lu, N.; Akinwande, D. *Acs Nano* 2013, 7, (6), 5446-5452.

(18) Yazyev, O. V. *Phys. Rev. Lett.* 2008, 101, (3), 037203.

(19) Cheng, Y. C.; Zhu, Z. Y.; Mi, W. B.; Guo, Z. B.; Schwingenschloegl, U. *Phys. Rev. B: Condens. Matter Mater. Phys.* 2013, 87, (10), 100401.

(20) Wang, C.; Wu, Q.; Ge, H. L.; Shang, T.; Jiang, J. Z. *Nanotechnology* 2012, 23, (7), 075704.

(21) Cao, C.; Wu, M.; Jiang, J.; Cheng, H. *Phys. Rev. B: Condens. Matter Mater. Phys.* 2010, 81, (20), 205424.

(22) Wang, S.; Ge, H.; Sun, S.; Zhang, J.; Liu, F.; Wen, X.; Yu, X.; Wang, L.; Zhang, Y.; Xu, H.; Neuefeind, J. C.; Qin, Z.; Chen, C.; Jin, C.; Li, Y.; He, D.; Zhao, Y. *J. Am. Chem. Soc.* 2015, 137, (14), 4815-4822.

(23) Wu, F.; Huang, C.; Wu, H.; Lee, C.; Deng, K.; Kan, E.; Jena, P. *Nano Lett.* 2015, 15, (12), 8277-8281.

(24) Perdew, J. P.; Burke, K.; Ernzerhof, M. *Phys. Rev. Lett.* 1996, 77, (18), 3865-3868.

(25) Kresse, G.; Furthmuller, J. *Phys. Rev. B: Condens. Matter Mater. Phys.* 1996, 54, (16), 11169-11186.

(26) BLOCHL, P. E. *Phys. Rev. B: Condens. Matter Mater. Phys.* 1994, 50, (24), 17953-17979.

(27) Dudarev, S. L.; Botton, G. A.; Savrasov, S. Y.; Humphreys, C. J.; Sutton, A. P. *Phys. Rev. B: Condens. Matter Mater. Phys.* 1998, 57, (3), 1505-1509.

(28) B. deB. Darwent, *National Bureau of Standards*, No. 31, Washington, DC, 1970.

(29) Peng, Q.; De, S. *Phys. Chem. Chem. Phys.* 2013, 15, (44), 19427-19437

(30) Topsakal, M.; Cahangirov, S.; Ciraci, S. *Appl. Phys.Lett.* 2010, 96, (9), 091912.

(31) Stoner, E. C. *Proc. R. Soc. Lond. Ser. A Math. Phys. Sci* 1938, 165, (A922), 0372-0414.

(32) Stoner, E. C. *Proc. R. Soc. Lond. Ser. A Math. Phys. Sci* 1939, 169, (938), 339-371.

(33) Pan, H.; Yi, J. B.; Shen, L.; Wu, R. Q.; Yang, J. H.; Lin, J. Y.; Feng, Y. P.; Ding, J.; Van, L. H.; Yin, J. H. *Phys. Rev. Lett.* 2007, 99, (12), 127201.

(34) Peng, H.; Xiang, H. J.; Wei, S.; Li, S.; Xia, J.; Li, J. *Phys. Rev. Lett.* 2009, 102, (1), 017201.

(35) Anderson, P. W. *Phys. Rev.* 1950, 79, (2), 350-356.

(36) Goodenough, J. B. *Phys. Rev.* 1955, 100, 564.

(37) Goodenough, J. B. *J. Phys. Chem. Solids* 1958, 6, 287.

(38) Kanamori, J. *J. Phys. Chem. Solids* 1959, 10, 87.

(39) Khomskii, D. I. in *Spin Electronics*, Ed. Ziese, M. and Thornton, M. J. Berlin, 2001; pp 89-116.

(40) Eda, G.; Fujita, T.; Yamaguchi, H.; Voiry, D.; Chen, M.; Chhowalla, M. *ACS Nano* 2012, 6(8), 7311-7317.

# Figures

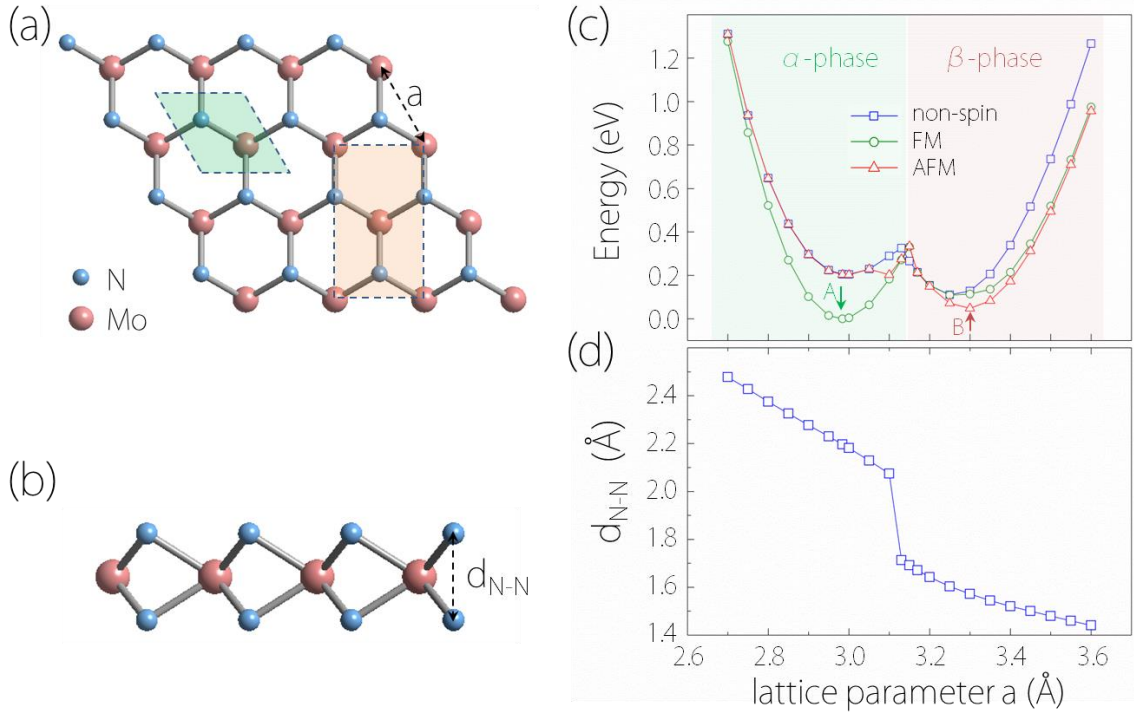

**Figure 1.** (a) Top and (b) side views of a 2D MoN$_2$ monolayer. In (a), the green shaded region indicates the primitive unit cell, and the red shaded region indicates the enlarged unit cell for the striped AFM configuration (for which the two Mo sites in the cell have opposite spin polarizations). $a$ is the lattice parameter. In (b), $d_{\text{N-N}}$ denotes the N-N distance. (c) System energies for the three types of magnetic configurations (non-spin polarization, FM, and AFM) versus the lattice parameter. A and B denote the two (local) energy minimum states in the $\alpha$-phase and the $\beta$-phase, respectively. All the energies are referenced to the energy of state A. (d) Variation of the N-N distance $d_{\text{N-N}}$ versus the lattice parameter (for the non-spin polarized case), showing an abrupt change at the phase transition. The curves for FM and AFM configurations are almost the same hence are not shown.

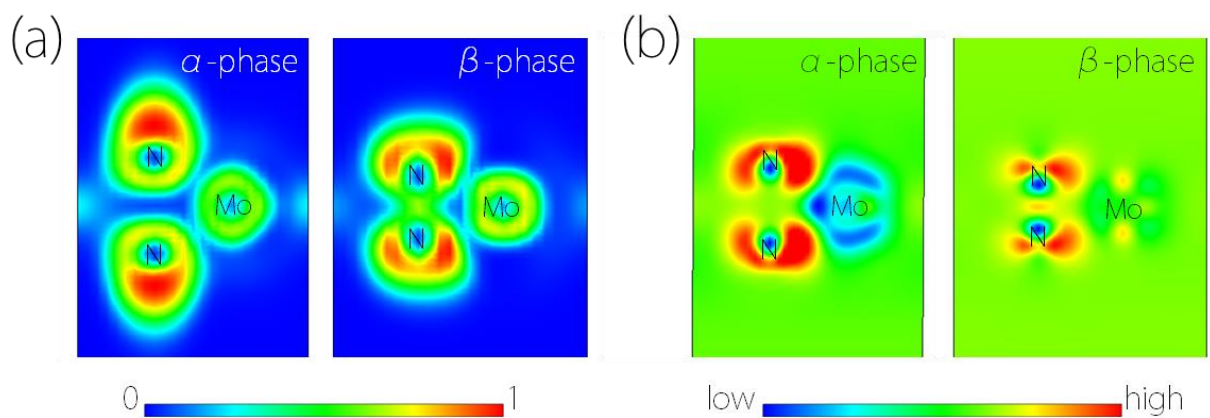

**Figure 2.** Maps of (a) the electron localization function (ELF) and (b) the deformation charge density for the $\alpha$-phase (of state A) and the $\beta$-phase (of state B). The plot is of the vertical plane cutting through one Mo atom and two N atoms. The results indicate a N-N non-bonding to bonding transformation across the isostructural phase transition.

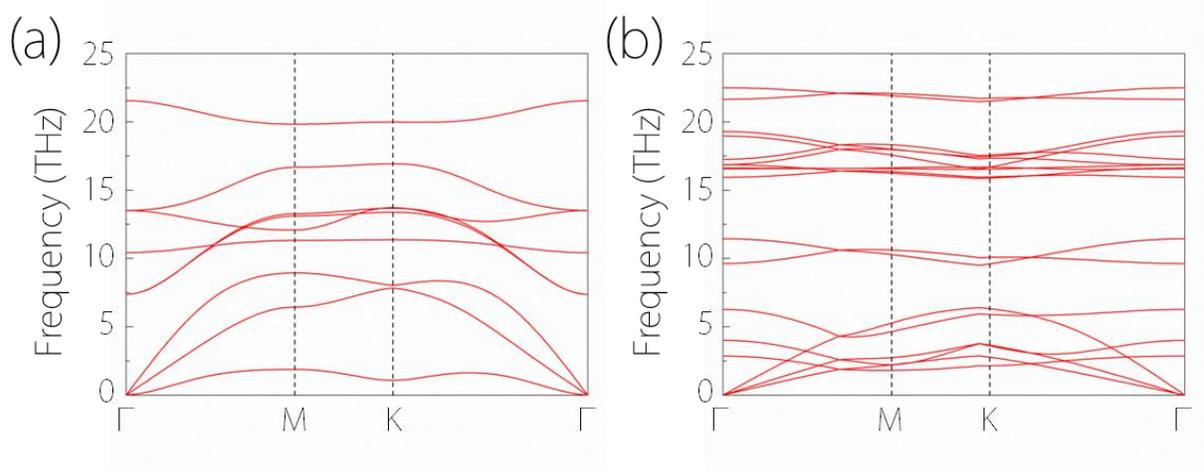

**Figure 3.** Phonon dispersions for the (a) state A in $\alpha$-phase and (b) state B in $\beta$-phase. A $2 \times 2$ supercell is used in the calculations.

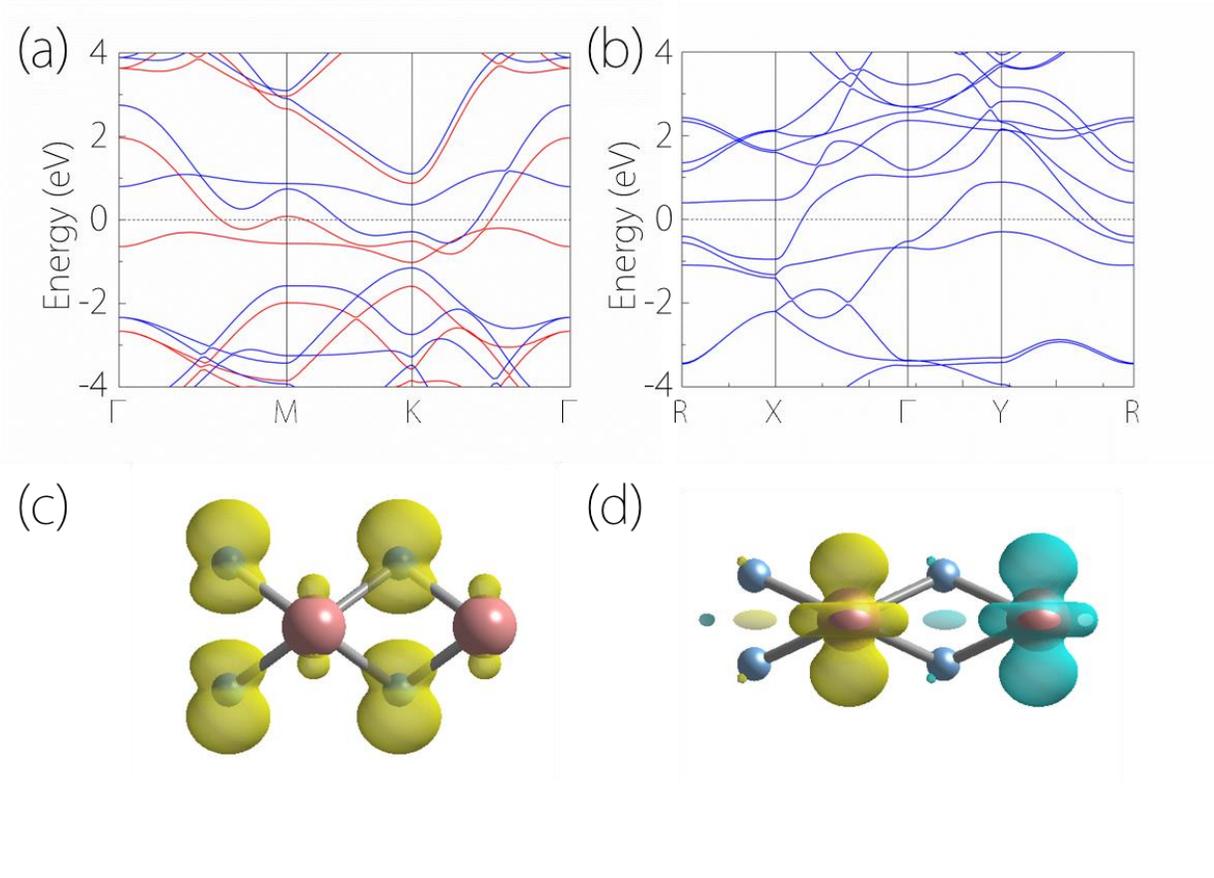

**Figure 4.** Electronic band structures for (a) the FM state A and (b) the AFM state B. In (a), the red and blue colors indicate the spin-up and spin-down bands. (c) and (d) show the spin polarization distribution (in the side view) for the state A and the state B, respectively. The yellow (cyan) color indicates a net spin-up (spin-down) polarization. In the FM state (c), the spin polarization is mainly on the N sties, whereas in the AFM state (d), the spin polarization is mainly on the Mo sites.

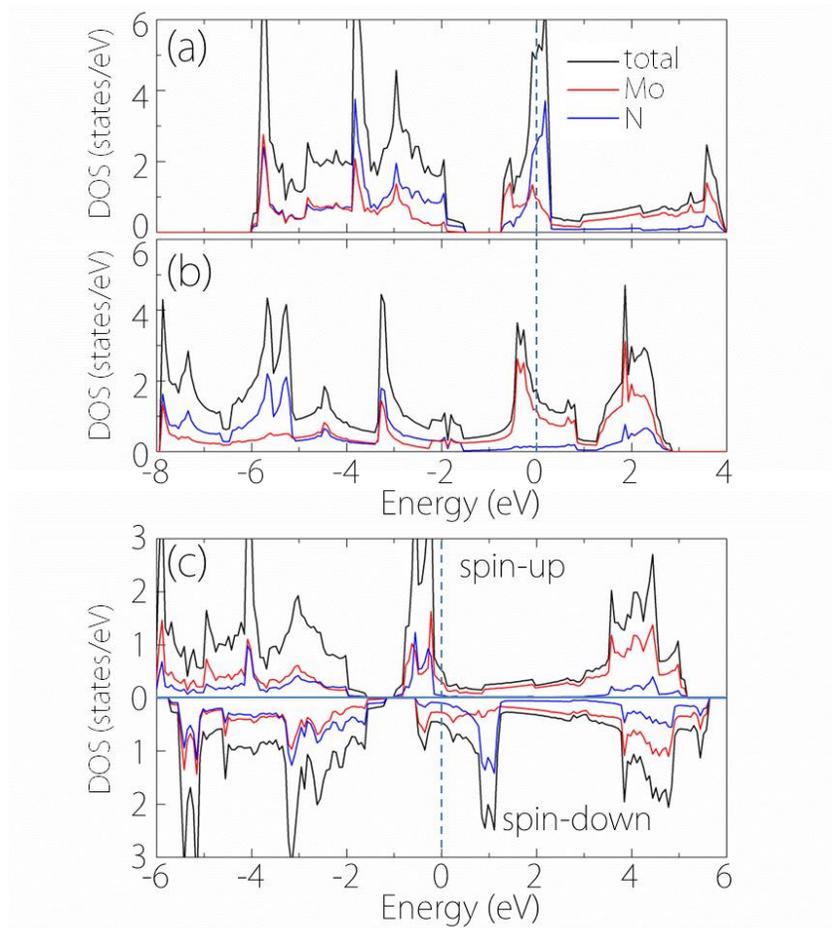

**Figure 5.** Density of states (DOS) corresponding to the structures of (a) state A and (b) state B but with non-spin polarization. (c) Spin-resolved DOS of the FM state A. In the figures, the blue and red curves are the projected DOS onto the N(2$p$) and Mo(4$d$) orbitals, respectively. The dashed lines indicate the locations of Fermi level (which is set as energy zero).